# A global colour mosaic of Mars from Mars Express HRSC high altitude observations

G. G. Michael[1,2], D. Tirsch[3], K.-D. Matz[3], W. Zuschneid[2], E. Hauber[3], K. Gwinner[3], S. H. G. Walter[2], R. Jaumann[2], T. Roatsch[3], F. Postberg[2], J. Liu[1]

[1]Center for Lunar and Planetary Sciences, CAS Institute of Geochemistry, Guiyang, 550002, China
[2]Planetary Sciences and Remote Sensing, Institute of Geological Sciences, Freie Universität Berlin, Berlin 12249, Germany
[3]Institute of Planetary Research, German Aerospace Center (DLR), Rutherfordstrasse 2, 12489 Berlin, Germany

### Highlights

- Use iterative method with high-altitude images to determine global brightness models in five colours
- Derive global high-altitude image mosaics at 2 km/pix for each of the HRSC colour filters
- Demonstrate use of high-altitude mosaic to reference full-resolution colour mosaic products

### Abstract

The ever-changing transparency of the Martian atmosphere hinders the determination of absolute surface colour from spacecraft images. While individual high-resolution images from low orbit reveal numerous colour details of the geology, the colour variation between images caused by scattering off atmospheric dust can easily be of greater magnitude. The construction of contiguous large-scale mosaics has thus required a strategy to suppress the influence of scattering, often a form of high-pass filtering, which limits their ability to convey colour variation information over distances greater than the dimensions of single images. Here we use a dedicated high altitude observation campaign with the Mars Express High Resolution Stereo Camera (HRSC) (Neukum and Jaumann, 2004; Jaumann et al., 2007), applying a novel iterative method to construct a globally self-consistent colour model. We apply the model to colour-reference a high-altitude mosaic incorporating long-range colour variation information. Using only the relative colour information internal to individual images, the influence of absolute image to image colour changes caused by scattering is minimised, while the model enables colour variations across image boundaries to be self-consistently reconstructed. The resulting mosaic shows a level of colour detail comparable to single images, while maintaining continuity of colour features over much greater distances, thereby increasing the utility of HRSC colour images in the tracing and analysis of martian surface structures.

## 1. Introduction

It is possible to make visual sense of a planetary surface image without knowing anything about the meaning of its pixel values beyond that they are proportional to the light received at the sensor. We might optimise it for display by merely adjusting the contrast and brightness, and at this point be satisfied that we can see all the information it contains. When the goal is to join two or more images together to appear as one larger image, many more elements have to come into consideration. The first are to do with geometry: for images to overlap precisely where they share coverage, it is necessary to understand where the spacecraft was at the moment of imaging in relation to the planet, the direction the camera was pointing, and how the image pixels map onto a potentially irregular planetary surface through the camera optics. Imperfections of these calculations cause distortion and misalignment of the images. Treating the geometry is a complex problem discussed elsewhere (Gwinner et al., 2016).

The second are to do with photometry: here we are concerned with relating the quantity of light accumulated by the image sensor to a physical property of the surface – its reflectivity. While, for a good quality Mars image, this is the dominant influence, there are numerous lesser ones that may be categorised across illumination, atmosphere, camera optics, sensor electronics, and data compression. The mechanisms of undesirable effects of the optics, electronics and compression can usually be understood and minimised by appropriate calibration and filtering. The illumination is more difficult to account for. An ideal camera image is not a direct representation of surface reflectivity, but rather a compound of reflectivity with surface orientation: surfaces more inclined towards the sun appear brighter, all else being equal. The relationship between apparent brightness and surface orientation is a function of illumination and observation angles, and also of the material and constituent particle size distribution. Human vision has a natural ability to deconvolve reflectivity from illumination, allowing us to see directly the three-dimensional form of an object simultaneously with its variations of colour. It is thus not always desirable to achieve a pure representation of reflectivity: we may learn as much or more from the form of the surface as from its reflectivity. Most planetary mosaics use images with the illumination as acquired, correcting only for its magnitude as it changes over the planet's curvature. This can produce visual peculiarities if images are acquired at different local times of day, with features appearing lit from two directions. Although in some circumstances it is possible to computationally alter the direction of illumination, this usually comes at the price of degraded image quality, and is therefore not routine. Instead, we plan and select, as far as possible, images taken at similar times of day, and where not possible, accept the visual peculiarities that occur.

The atmosphere, however, is the main difficulty for surface photometry of Mars, primarily because its dust content varies through the seasons in a way which is not precisely predictable. An image captured through dusty atmosphere becomes both brighter and reduced in contrast as light is reflected and scattered by the dust particles. To have a chance of compensating the effect using a physical model it is necessary to make an accurate estimate of the dust load. Most successful in this respect is the so-called shadow method (Kirk et al., 2001; Hoekzema et al., 2011; Shaheen et al., 2022), which uses the difference in brightness of similar, closely-neighbouring shadowed and sunlit areas to estimate the atmospheric optical depth. Nevertheless, the method has not been systematically applied for HRSC because it depends on an informed selection of sample areas in each image and, more importantly, good shadow areas occur only in images where the topography and illumination permit. The pixel size of HRSC high-altitude images of Mars (0.5—1 km), in particular, is generally too large to find complete pixels in shadow, so we do not attempt to apply the method in this work.

Previously, we had been addressing the systematic correction of HRSC images for atmospheric dust for the purpose of mosaicking not with a physical model, but instead by referencing each image to a known standard (Michael et al., 2016) – the Mars Global Surveyor Thermal Emission Spectrometer (TES) albedo map (Christensen et al., 2001). The absolute brightness of each image was scaled to an interpolation of the reference and thus forced to be consistent across image boundaries. The image contrast was restored by a visual adjustment of each image. This works well for the highest resolution panchromatic images, but could not be extended to colour images since there are no comparable albedo references for the wavelength passbands of the camera's colour filters. Instead, we had to restrict the colour mosaic to showing only local colour variation, i.e. that seen within a single image, and use the global average colour to enforce continuity across image boundaries. This allowed high resolution colour details to be studied over a wide area, but the 'zoomed out' view was near-monochromatic, albeit with the familiar Mars average red hue (see Gwinner et al. (2016), Fig 24, third panel).

Between 2015 and 2017, several high-altitude observations were acquired for calibration purposes which were also remarkable for the quality of surface colour variation that was captured, extending over far greater distances than the standard observations (Mars Express is in a highly elliptical orbit around the planet where its altitude above the surface varies between about 250 and 10 000 km. Imaging is normally carried out close to the surface to obtain the highest pixel resolution.) Global coverage with this type of observation offered the potential to improve on the current colour processing by increasing the distance range of retained local colour information. A dedicated high-altitude observation campaign for this purpose was started in 2019.

The new images could be joined together following the previous scheme to construct a colour mosaic, retaining the local colour variation over the range of a single image, but now with a typical surface footprint width of 2500 km instead of 50 km. However, since the set of images covers the whole globe with substantial overlaps, it turned out to be possible to do even better than this, by first using them iteratively to compute a global surface colour model – a set of global brightness models, one for each colour filter, as described in the following section.

## 2. Material and methods

### 2.1 Data and software

The source data are Mars Express HRSC radiometrically calibrated, map-projected VICAR images (a format originally developed at NASA JPL for multi-dimensional imaging data), with sub-spacecraft-point-centred orthographic now being the standard projection for newly acquired high-altitude images. Reprojection, cropping and merging operations were carried out with the Open-Source Geospatial Foundation's GDAL (Geospatial Data Abstraction Library) and image inspection with their QGIS geographic information system. Other operations were made with in-house routines coded in IDL (a programming language currently owned by NV5 Geospatial Solutions, Inc.), also making use of the NAIF SPICE Icy toolkit – NASA's Navigation and Ancillary Information Facility's observation geometry information system (Acton et al., 2018).

### 2.2 Image acquisition, selection and sequencing

Images are acquired at altitudes between 4 000 and 10 500 km above the surface by carrying out a slew manoeuvre: the spacecraft rotates to sweep the field of view of the camera's imaging sensor lines over the surface of the planet (in contrast to low-altitude observations, where the orbital velocity moves the field of view over the surface). Such observations cover a strip centred on the sub-spacecraft point, encompassing roughly one third of the spacecraft-facing disc of the planet, and requiring approximately 3 minutes each to complete. During some later observations a second sweep was made, offset from the first, thereby increasing the coverage (after an interval of about 25 minutes to reverse the rotation, repoint and stabilise). At the time of writing, 320 images had been acquired with pixel scales >200 m, of which 92, selected for quality of surface imaging with dust levels as low as possible are used in this work.

The intersection of the observation footprints on a global map is rather complex because they are variously oriented and frequently overlap the poles and dateline, making an optimal selection and sequencing of images difficult. The problem was simplified by dividing the surface into six equivalent orthographically-projected faces centred on the two poles and equatorial points at 0°, 90°, 180°, 270°E, each extending to 2800 km radius. For a Mars nominal projection radius of 3396 km, this is sufficient to cover the globe without gaps while avoiding extreme projection distortion. A selection and sequencing were made for each, discarding lower quality images, or pushing them down in the sequence where there were no alternatives, trimming out significant regions of cloud, and choosing polar images with minimal seasonal ice cover. Finally, the six sequences were combined for the whole globe to ensure consistency over the face boundaries.

### 2.3 Preprocessing

The high-altitude images are map-projected using a sub-spacecraft-point centred orthographic projection, chosen because the image shape is as close as possible to that of the raw sensor image, without singular points, and of equivalent behaviour at all coordinates. The Lambert correction of the magnitude of illumination over the planet's curvature is carried out in this projection, with those pixels having a solar incidence angle greater than 85° discarded, as well as any pixels which were already saturated. We do not attempt to apply higher order correction models since the observation/illumination angles for which they provide most benefit also coincide with longer atmospheric paths, which we already exclude.

Images are next cropped against a set of manually constructed polygons to exclude obvious regions of cloud, ice cover or data defects, and clipped to a radius of 2200 km from the sub-spacecraft point. This value was chosen as a trade-off between maximising coverage and excluding parts of the image where the longer atmospheric path causes a significant divergence in scattering behaviour: a raw colour image becomes notably bluer towards the limb (Figure 1a). Coverage is further restricted to those surface pixels which are present in all colour filter images (for HRSC, these have similar, but not identical surface footprints).

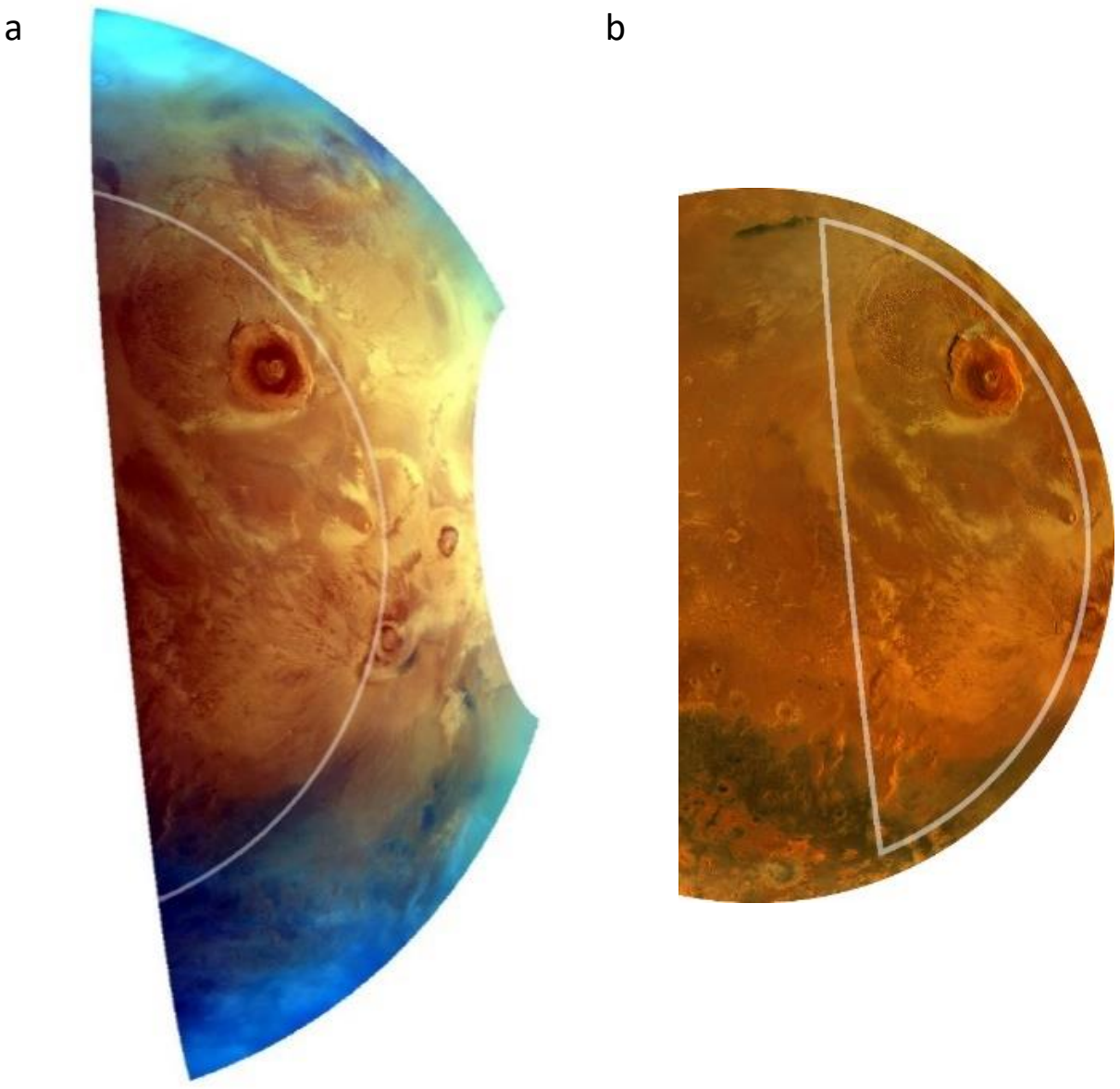

Figure 1. a) HRSC image hl678_0000, orthographic projection centred on sub-spacecraft point (off-nadir pointing). Raw RGB composite showing the colour shift towards blue approaching the limb. Partial circle indicates 2200 km radius of sub-spacecraft point used to clip out limb. b) Completed RGB mosaic shown in same projection, showing carry-over of detail from utilized image region and continuity of detail and colour with neighbouring areas.

### 2.4 Constructing a surface brightness model

The aim at this stage is to construct a low-resolution model which captures the global brightness variation over distances greater than covered by individual images, which have a swath width of up to 2500 km. The following is an outline of the procedure, carried out independently for images acquired through each filter:

1. Preprocess images: project to sub-spacecraft-point centred orthographic projection; apply Lambert correction.

2. Generate a Fibonacci lattice: a set of points near-evenly distributed over the globe (Swinbank and James Purser, 2006), that is used to represent the brightness model (Figure 2). It is a style of representation of a global image with the advantageous property that there are no singular points or regions of extreme distortion. It is not commonly used in planetary imaging because its pixels are of irregular shape, making many typical operations on their values more computationally intensive (further details in section 2.5).

3. Initialise each point of the lattice to have a relative brightness of 1.0, representing the global mean for the filter.

4. For each image, at a pixel-scale reduced to roughly half the lattice spacing for speed of processing, transform the relevant part of the lattice to the image projection at the same resolution, and multiply the image reflectance values by $m_L/m_I$ where $m_I$ is the mean of the image reflectance pixels, and $m_L$ is the mean of the image-projected lattice pixels over the same area. On the first occasion, $m_L$ = 1.0, so that we are simply scaling the image values relative to their mean.

5. Mosaic the image into the lattice, maintaining continuity at the image boundary (see section 2.6).

6. Iterate from step 4 using the current state of the brightness model lattice as the new starting point until convergence is achieved. Ideally, convergence would mean there is no further change to the pixel values in the lattice. In practice, because image overlaps are not in perfect agreement as a consequence of, for example, the changing presence of clouds, seasonal frost, atmospheric transparency or other surface features, the solution instead reaches a minimally oscillating state. By comparing the lattice values at different stages, we found no further systematic improvement to the solution to occur after 12-15 iterations of the complete image set.

Note that the mean brightness of all the points in the lattice remains 1.0 throughout. The effect of the algorithm is to allow the surface brightness to be redistributed over the globe while maintaining this constraint. In early iterations, disagreements at image overlaps cause the 'area-integrated brightness' to move from darker to brighter regions. Thus, an image containing a polar cap, much brighter than the average terrain, after the first pass of step 5, will appear in the lattice with the cap being the brightest part (although less bright than its true value), and the surrounding terrain being darker than its true value. An adjacent image not including the cap, but which overlaps some of this surrounding terrain will disagree with the state of the model in the overlap region and push those values back upwards. On the next iteration, the mean value of the region covered by the polar cap image will thus be greater than 1.0, and will continue to rise in subsequent iterations until the overlaps are in agreement, the model eventually converging to a self-consistent state. At this point, the polar cap should also have achieved its true brightness relative to other terrains over the whole of the globe, the correct share of the total global brightness having been shifted into the region of the polar cap image.

The procedure disregards the absolute brightness values of the source images, relying only on relative brightness differences internal to individual images. This sidesteps the absolute calibration problem caused by the atmospheric dust, although a potential second order loss of contrast remains. Because the images were already selected for dust levels as low as possible, the problem is minimised, but the consequence of any dusty areas that do remain is to suppress somewhat the colour variation in the region of that image.

The brightness model lattice for each of the camera's colour filters can be interpolated to a standard map projection and used analogously to the TES albedo in previous work for brightness referencing a full resolution mosaic. Indeed, the same can be done for the panchromatic filter, removing the need to use the TES reference at all. This is advantageous in that some Mars albedo features have changed in the time between the operation of TES and HRSC, and inconsistencies between the reference and images cause artefacts in the mosaic, often appearing as a kind of fringing where a high contrast albedo boundary has moved on the surface. Also important is that the wavelength sensitivity of TES, 0.3–2.9 µm (Christensen et al., 2001), is substantially broader than the HRSC panchromatic filter's passband of 0.58–0.77 µm (Neukum and Jaumann, 2004). While the albedos over the different ranges correlate to a large degree, it is obviously preferable to have a reference matched to the filter.

### 2.5 Fibonacci lattice

In contrast to a map projection, such as the often used *simple cylindrical* where the polar areas of the globe are greatly distorted, a Fibonacci lattice is a truly spherical representation of a global image (Figure 2). While mathematically there is no perfect solution to the problem of placing equally spaced points on a sphere, excepting a few special cases with small numbers of points, the Fibonacci lattice is a very close approximation. Each point has a similar adjacent area – the area closer to the point than to any other – and there is no preferred region on the surface where the average adjacent area is greater or lesser. The iteration of the brightness model requires that each pixel of the representation relates to its neighbours similarly for homogeneous numerical behaviour towards convergence (we used this representation in a different planetary context for similar reasons (Liu et al., 2020)). In this work we use a lattice of 40 001 points which corresponds to a roughly 60 km spacing.

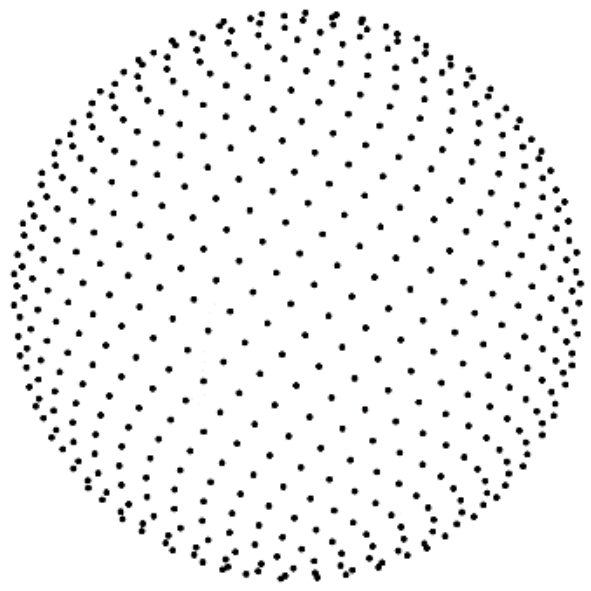

Figure 2. Fibonnacci lattice. Points are near-evenly distributed over the globe rather than over a projected plane, providing a means to represent a spherical image.

### 2.6 Enforcing continuity at image boundaries

Images are merged into the mosaic using a novel method to enforce continuity at the boundary between the image and the current state of the mosaic. It is occasionally suggested that this can be achieved by means of feathering alone, a technique where the transparency of an added image is varied from 0.0 to 1.0 over a small distance from its boundary. This has the effect of softening hard edges, but is only satisfactory if the absolute brightnesses of the images are already in good agreement. Sometimes it is possible to use observations constrained to a short time interval to ensure consistent atmospheric conditions. A sequence of colour images obtained over a few days in 1976 during the Viking II orbital approach allowed the long-range variation to be retained with some smoothing (Soderblom et al., 1978), and the four-week Mars Orbiter Camera (MOC) geodesy campaign opted to use a manually adjusted low-resolution mosaic acquired over the course of a day to adjust the long-range shading of high resolution images (Caplinger and Malin, 2001). A different approach is to average over many images of the same area to reduce the effects of atmospheric variation (Robbins, 2020). More commonly, an always robust solution is to apply a high pass filter to retain high-resolution detail (Kirk et al., 2000), as done for our previous colour processing (Michael et al., 2016), but long-range variations are then lost. The new method described here retains high-resolution detail similarly to a high-pass filter, but differs in that it retains progressively longer-range information towards the centre of each image. It also has the property that image overlaps are constructive of the longer-range detail so that when used in combination with the iteratively obtained brightness model, it becomes possible for the whole of the long-range variation to be restored.

The image is first divided up into square cells of different sizes in a tree-structured configuration. Pixels in contact with the boundary constitute cells in themselves of size 1. Pixels in contact with size 1 cells are grouped into cells of 2x2 pixels (size 2) as far as possible, with the remainder being additional size 1 cells. Pixels in contact with size 2 cells are grouped into cells of 4x4 pixels, with the remainder being size 2 cells and so on, doubling in dimension until the area of the image is completely covered. As a result, the largest cells are at the centre of the image with a gradation of diminishing in size cells towards the boundary in all directions (Figure 3a,b).

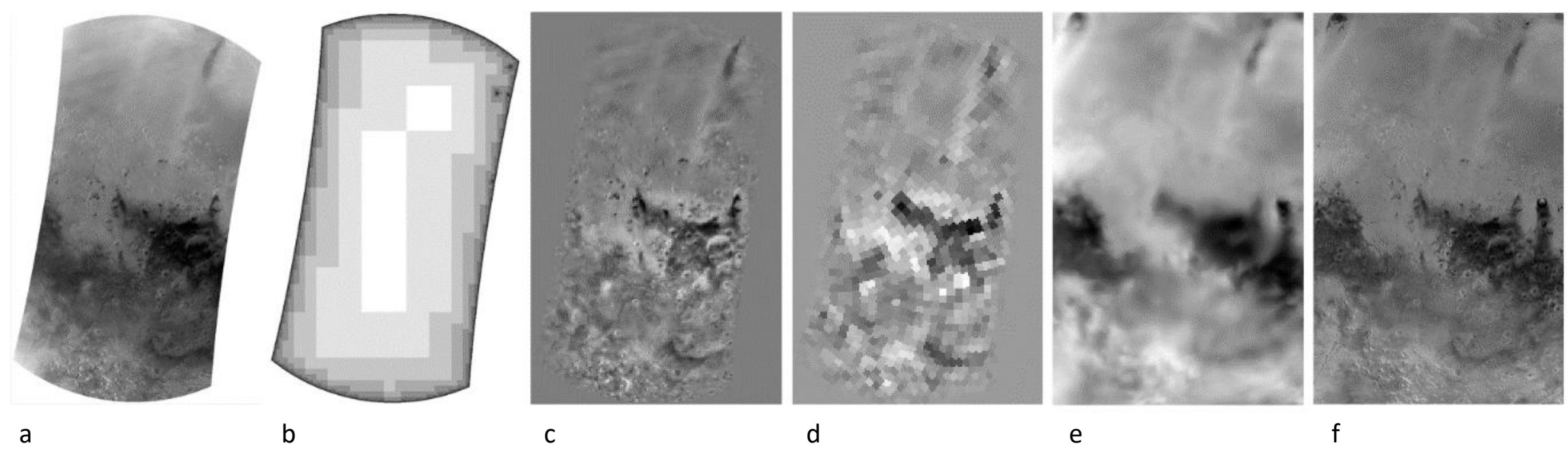

Figure 3. Red filter image hk725_0001 a) after preprocessing in sub-spacecraft-point centred orthographic projection, b) its tree-structured cell configuration: single pixel square cells at the boundary repeatedly double in dimension toward the centre of the image (cells coloured by size; note internal image holes upper-right, caused by excluded saturated pixels), c) after cell referencing to interpolated initial state of brightness model (mid-grey represents initial value, 1.0: note edge continuity with this value), d) after downsampling to Fibonnacci lattice pixels to insert into brightness model, e) same region from final state of brightness model after 8 iterations (upsampled), f) region from final mosaic in same projection, after referencing of all images to the brightness model. Note that short-range information is carried over from the original image, and long-range information from the brightness model.

For each cell, we find the ratio between the total of the pixel values in the new image and those of the corresponding area of the current state of the mosaic, and make a bilinear interpolation of the ratios from the centre of each cell over the whole area of the image. Finally, the new image is divided by the interpolation and substituted into the mosaic. This forces exact agreement of size 1 cells; larger cells can differ in details smaller in scale than the cell, but always agree on the mean. The procedure removes any chequerboard effect produced by images of different absolute brightness levels as well as providing edge transitions without steps. Larger cells allow longer range information to be transferred into the mosaic, making it desirable to have them as large as possible. However, to make an interpolation scheme that enforces agreement at the boundary requires them small there, and to avoid interpolation artefacts – in particular, significant interpolation parallel to the boundary – there should also be a gradation of scales towards the boundary. The chosen tree-structured schema is not the only possibility to satisfy these conditions, but proves to be an effective choice. In our previous work (Michael et al., 2016) we had used a fixed cell size – which led to a trade-off between better edge agreement and maximising the length-scale of information transfer. Using this configuration allows us to improve both ends of the trade-off. Each image added into the mosaic thus appears completely seamlessly in relation to what is already there, both in terms of its edge – the pixels are identical at the boundary – but also its general brightness: the mean of every cell remains invariant.

The procedure is applied both during the iterative construction of the surface brightness model and during the subsequent high-resolution brightness referencing (next section).

### 2.7 Mosaic assembly, colour calibration and high-resolution brightness referencing

Each filter's images are processed at full resolution into six orthographic face projection mosaics, likewise chosen to avoid singular points or regions of extreme distortion. Each mosaic is initialised with an upsampled projection of the appropriate filter brightness model, and the images added one by one enforcing both brightness consistency and boundary continuity as described in section 2.6 (Figure 4). The purpose at this stage is to add in the detail that is not already in the brightness model lattice. The tree-structured cells near the centre of an image are larger than the lattice spacing, and therefore restore local variation over a range greater than that. Assuming the lattice model solution was good with respect to long-range features, the combination theoretically enables a restoration of image information at all scales. Near the edge of an image, the cells are smaller than the lattice spacing: here there is not a complete restoration. However, provided there are sufficient image overlaps, a later large cell can add more information, while a small cell never detracts from what is already present. Thus, the only areas where there is incomplete colour information are those which have not been covered by cells larger than the lattice spacing: most obviously, those close to the remaining few observation gaps.

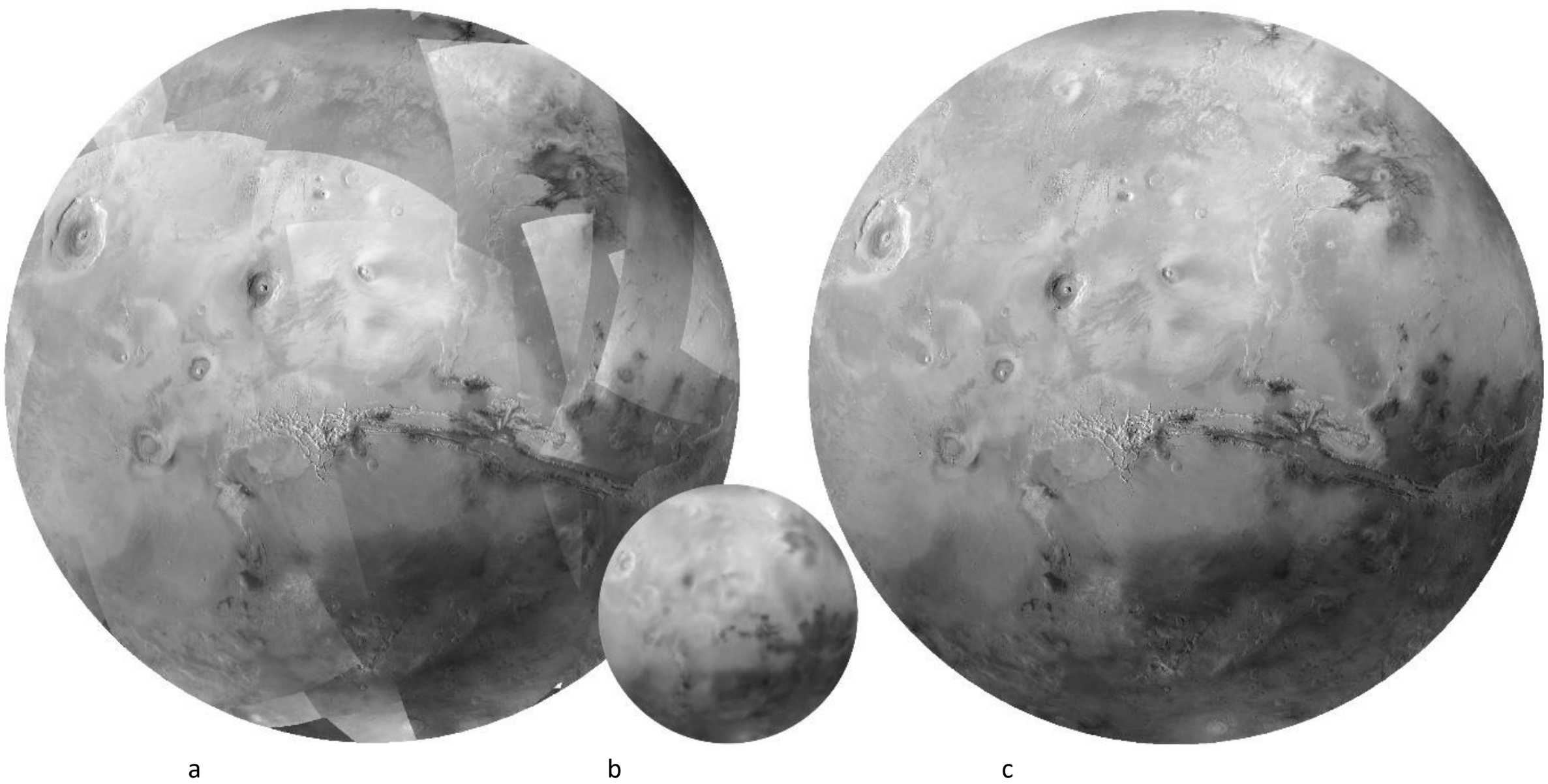

Figure 4. Orthographic face extending to 2800 km radius centred on 270°E 0°N, in panchromatic filter: a) pre-processed images, b) brightness model, c) final mosaic.

Since the five filter brightness models and mosaics are all generated on a relative scale (recalling that the mean is held at 1.0), colour calibration is carried out by choosing an area of good quality colour from an arbitrary single multi-filter image, calculating the mean colour reflectance values from it, and doing likewise for the same area of the colour model (the colour model being the set of brightness models for all filters). The ratio for each filter may be used to rescale from relative brightness to actual colour reflectivity. This yields a colour quality in that region close to the calibration of the original image. Provided that the colour model is a good solution, the calibration area should extrapolate properly to the rest of the globe. Qualitatively, it appears to hold well through the majority of the non-ice-covered terrain of the planet, only becoming somewhat strained at the north polar cap. This is perhaps unsurprising, since the surface colour of most of the planet's terrain varies rather modestly in absolute terms, but ice features are strongly different. Thus, where numerical errors do arise, they are expected to show strongest in the vicinity of the ice caps. Compounded with this is the additional difficulty of obtaining well illuminated polar observations near minimum ice cover because of the additional constraints on time of year and spacecraft orbital position, which means that coverage is still not complete for a part of the north cap.

The face mosaics are easily assembled and transformed into other projections without introducing artefacts, and the red, green, and blue filter mosaics – or other combinations or ratios – may be combined into a colour composite (Figure 5).

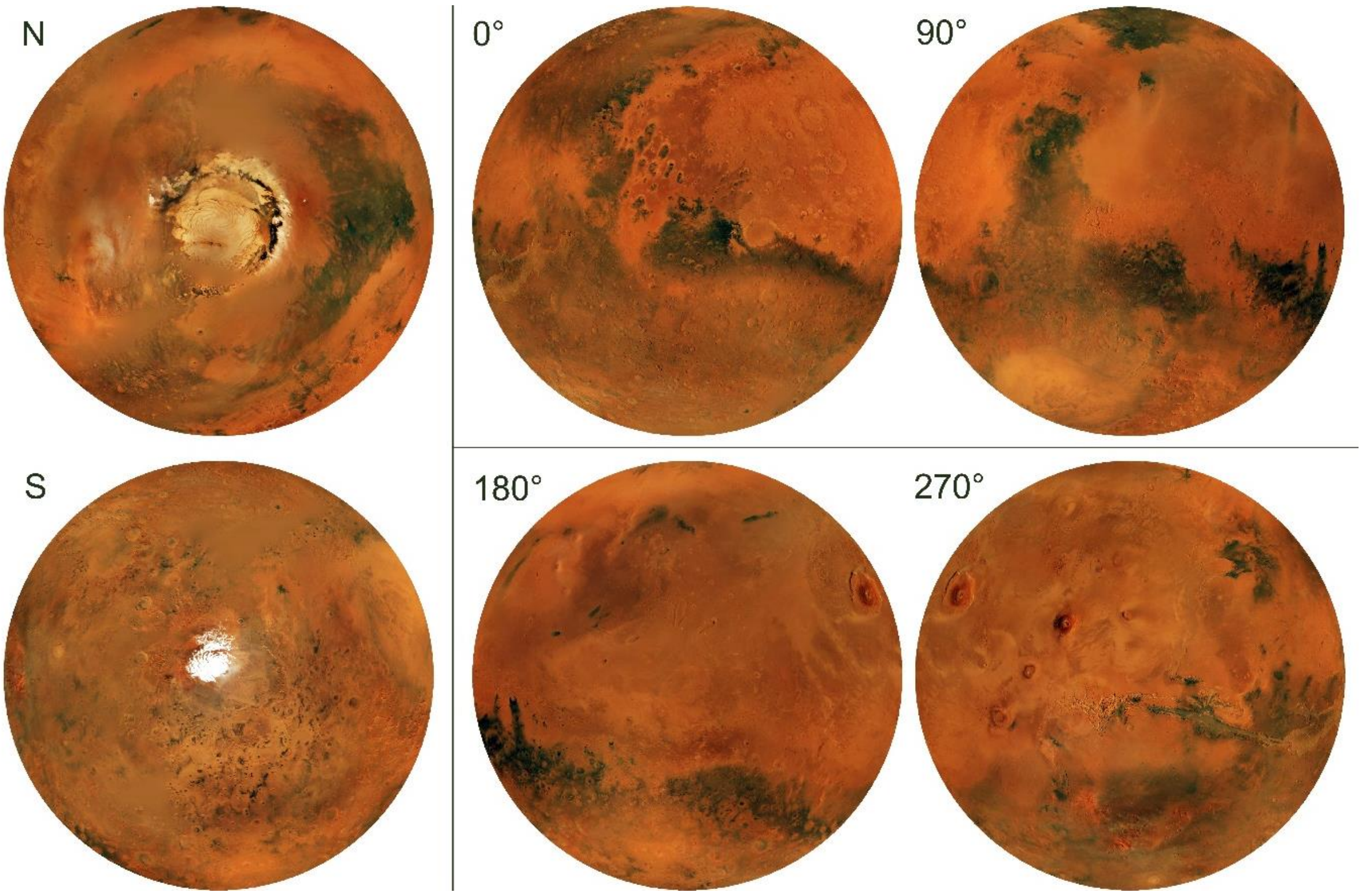

Figure 5. Six orthographic faces centred on the two poles and equatorial points at 0°, 90°, 180°, 270°E, each extending to 2800 km radius making up the global RGB high-altitude mosaic. Representation avoids singular points and regions of extreme distortion.

While the absolute variation of colour is small over most of the mosaic, the variations which are present are maintained with good fidelity in the completed mosaic and may be exaggerated for visualisation with a colour stretch (Figure 6). Note that, because the polar ice caps are quite distant in colour space from the normal terrain, it is difficult to find a stretch which simultaneously enhances the differences in normal terrain without pushing the ice regions beyond saturation.

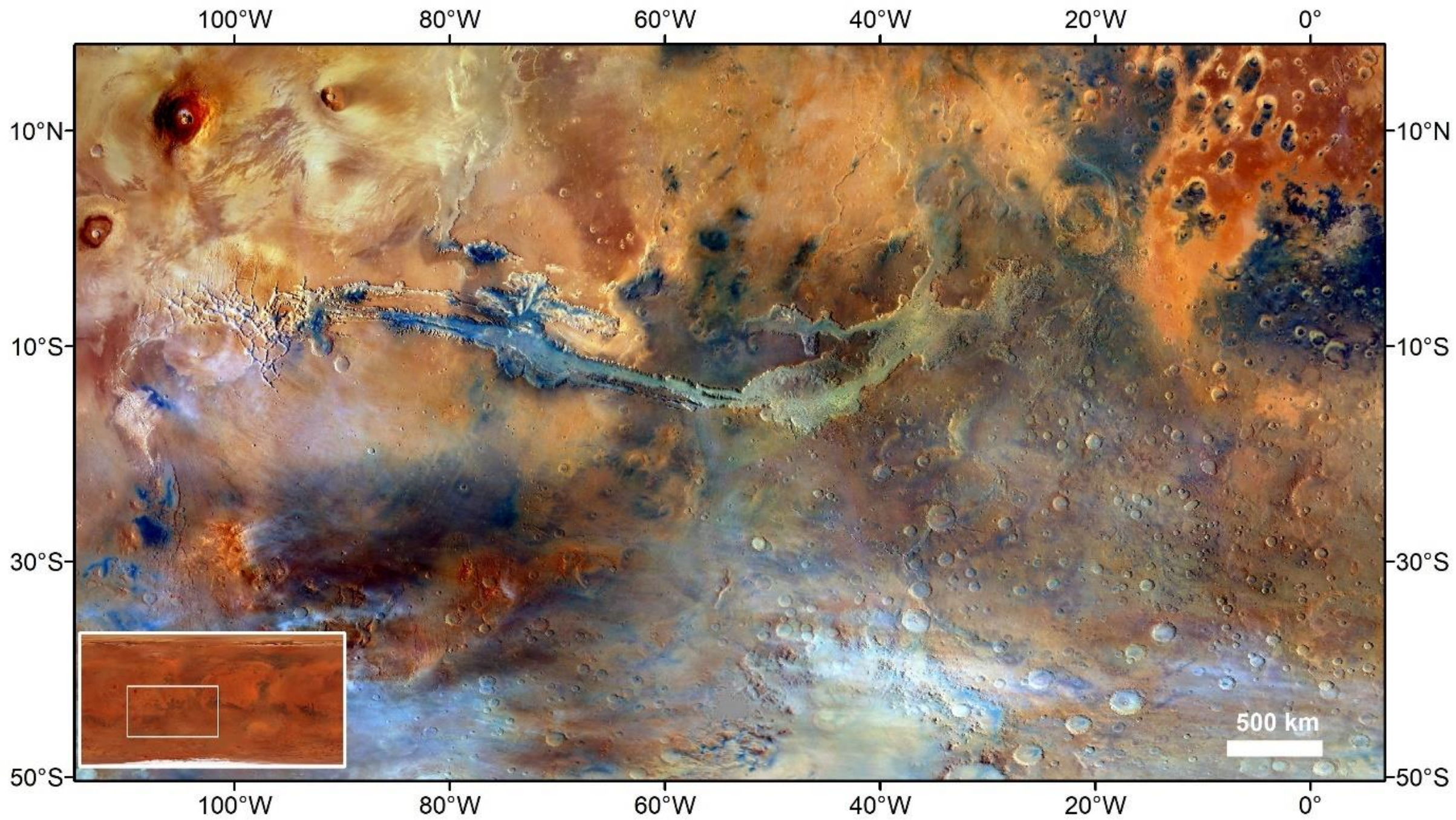

Figure 6. Valles Marineris – Argyre region with exaggerated local colour stretch, enlarged from global RGB high-altitude mosaic (inset). Notably distinct colour features are the Tharsis lava flows, an arcuate boundary south of Noctis Labyrinthus, the light region at the north of Claritas Fossae, the dark material in Valles Marineris together with its lighter fog, the bright etched unit (Hynek and Di Achille, 2017) in Meridiani Planum, and the cloud formations to the south.

Brightness referencing of high-resolution images acquired from low altitude is essentially a repetition of the same procedure at smaller scale, where the global colour mosaic becomes the colour brightness reference for the smaller coverage images, allowing the processing of a full-resolution continuous colour mosaic (Figure 7).

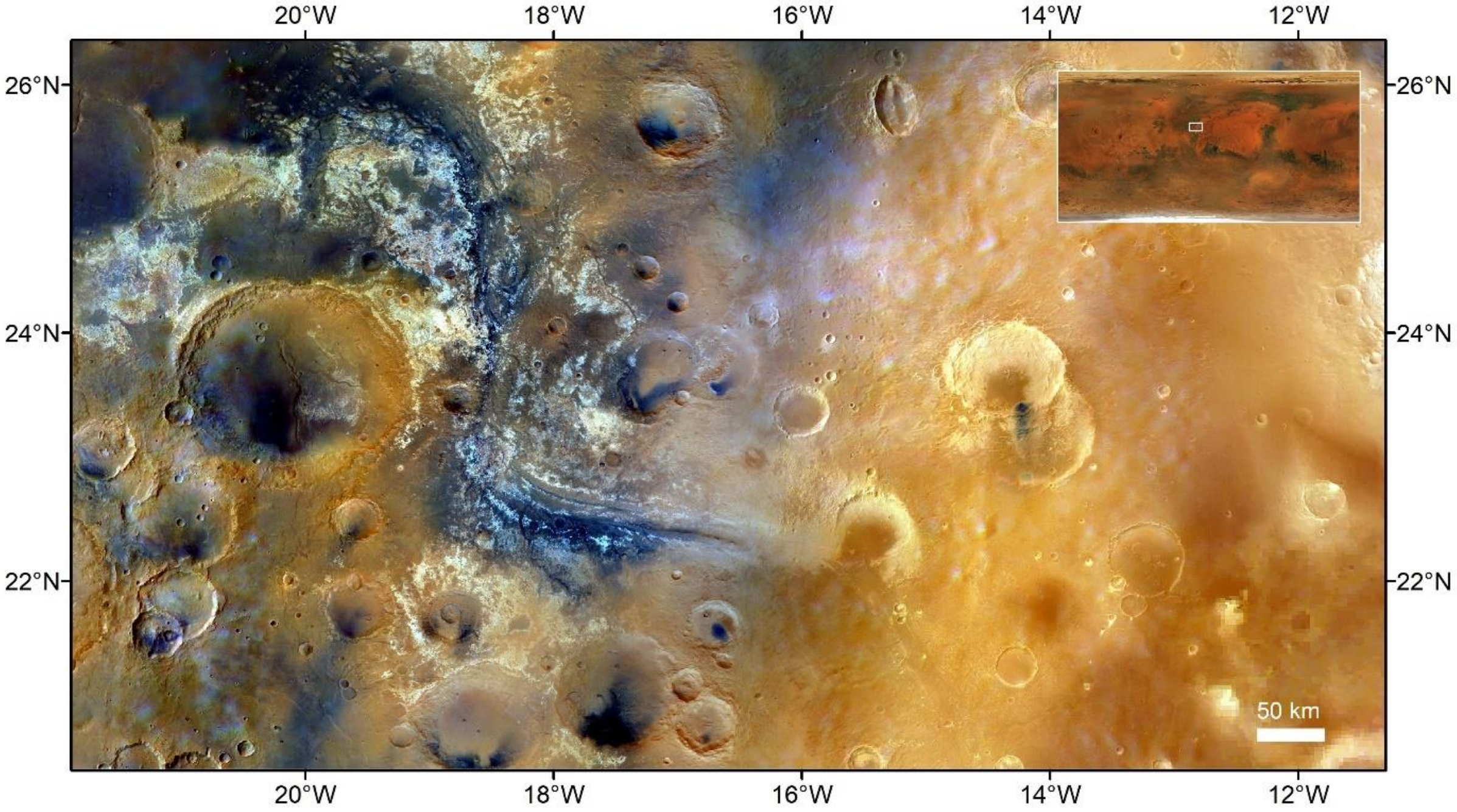

Figure 7. High resolution colour mosaic of Mawrth Vallis region constructed from 13 sets of red, green, and blue filter images, each brightness referenced to the respective colour high-altitude mosaic. Note that the colours separate at least four distinct surface materials in addition to the cloud formations.

## 3. Discussion

Both the iteration of the colour model and the construction of the mosaic are designed such that the addition of a new image leaves the average brightness of the globe invariant. A brightness model starts out with the uniform value 1.0 everywhere, and evolves to a brightness-resolved state which conserves that average value. Non-ideal images (those which contain non-homogeneous disagreements in overlap areas, e.g. caused by a cloud) cause local oscillations of the model solution, resulting in an absolute value in that area which is uncertain by the magnitude of the disagreement. This uncertainty limits the accuracy of propagation of the overlap agreements. Since the full set of images make up an interlinked network covering the globe, the influence of a single image defect elsewhere is not obvious, but it is evident that since the solution stabilises, the effect must diminish rather than grow with distance through the network. It may be understood that alternative routes through the network with either lesser or randomly opposing error have the effect of suppressing its effect further away.

A part of the problem in evaluating a colour mosaic like this one is that there is no reference dataset with which to make a meaningful comparison. Of course, there are localised measurements from many instruments, but for the wavelength bands of interest, these suffer from identical intercomparability difficulties to individual HRSC observations. A possible test of success in the mosaic construction may be to ask: 1) does the mosaic faithfully reproduce the colour variation seen in single images before mosaicking? (see Figure 1b) 2) Do the colour properties of the mosaic vary continuously and plausibly over image boundaries? (see Figures Figure 4,Figure 5, Figure 6) While we believe we can answer yes to both of these, there remains a third question, 3) how precisely are different regions of the globe positioned in colour space relative to one another? Despite visually consistent transitions across boundaries, there is the potential for drift in relative position over greater distances. Indeed, we know this occurs because of the remaining coverage gaps which leave those locations tied to the relative value of 1.0. Without coverage gaps and with ideal surface images, the colour model would converge on the correct values. For now, it appears to be the coverage gaps which are the main source of distortion of the model. When the gaps are filled, which is expected before the end of the extended mission, we expect that the absolute offsets should shrink, leaving the non-ideal properties of the surface images as the main source of error: the presence of clouds, seasonal frost, fogs, etc which cause inhomogeneous disagreements between overlapping images.

Test (1) refers to what we would consider to be the preservation of short-range features. In practice, wherever the interpolation cell size exceeds the colour model cell size we expect all shorter-range information to be transferred into the mosaic. Where there are image overlaps, the information is derived from both images, so that whichever provides the largest interpolation cell is the one that counts in this respect. Adjacent to image gaps or other areas with poor overlap, there may be cells which are smaller than the colour model cells implying some short-range losses. Test (3) obviously relates to the retention of long-range variations. Visually, in comparison with any high-pass filtered mosaic, it is apparent that they are substantially retained. We have to appeal to the mathematics to argue that in the case of complete coverage and near-ideal images, there is no constraint on how much long-range information could be captured.

There remain a few patches which have not yet been covered in the high-altitude campaign after accounting for illumination, view-angle, and ice or cloud cover constraints (indicated for the panchromatic mosaic in Figure 8a and also seen, particularly around north polar cap, in Figure 5). These are seen as featureless areas in the mosaic rather than obvious holes: because of the algorithm of mosaic construction, they remain continuous with their vicinity. Being continuous, but at the wrong absolute value causes a systematic distortion of the absolute values in their vicinity. It does not affect the short-range (high-resolution) detail, which remains true relative to the local average colour, but manifests in the long-range colour differences, so that the regional average colour may be suppressed in its relationship both to the global average and to neighbouring regions. The extent of the distortion depends on what was missing in the gaps: if, for example, it was typical red Mars terrain the difference may be small since the average value was already assumed. If it was an area of the north polar ice cap, the resulting distortion of the colour model is much greater. There are two gaps in the vicinity of the north polar cap, one of which cuts the cap and forces it to be continuous in colour with it (Figure 5). Since the colour model is initialised to value 1.0 – representing the average colour of the planet – the gaps are much less bright than the polar cap. The net result is to pull down the brightness of the cap in the model solution. Other gaps are less noticeable in their effect, but it should be noted that they also pin the value of the model solution to 1.0 where they occur. This likely restricts the extrema of brightness and darkness of the model solution.

### 3.1 Alternative processing using TES albedo map to seed initial model state

As an experiment to understand the pinning effect of coverage gaps, we tried seeding the initial state of each filter's brightness model with a normalised TES albedo map. The gaps in the mosaic thus end up pinned to the values of those areas in TES. The result is more TES-like, with greater contrast between the south and the lighter provinces of the northern hemi-sphere (Figure 8). While likely closer to a solution without HRSC gaps, it has a somewhat smudged look in places, suggesting some disagreement between the datasets about the value or location of the darkest areas, either because of the spectral difference or temporal changes. If the coverage were complete, the brightness model solutions would be expected to be independent of the initial state.

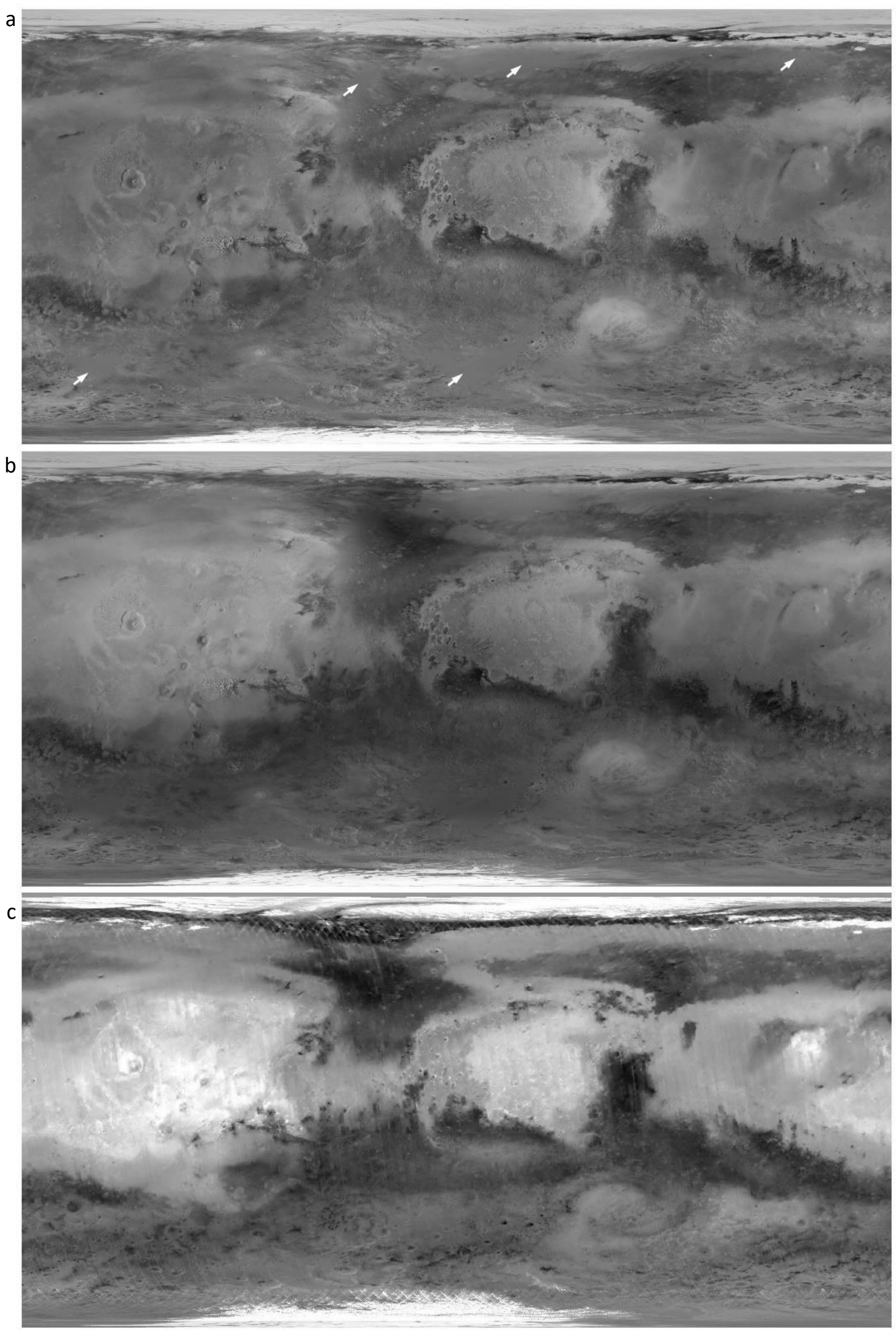

Figure 8. In equicylindrical projection with central zero meridian: a) HRSC global panchromatic high-altitude mosaic, b) alternative processing based on brightness model seeded with normalised TES albedo, c) TES albedo map (Christensen et al., 2001). White arrows indicate areas of no coverage (fewer than for colour channels).

## 4. Conclusion

An iterative approach to constructing a brightness model, together with careful boundary constraints on merging images into it, offers the possibility to derive a global model of relative brightness using image overlaps without access to estimates of atmospheric optical depth. We used the models to brightness-reference five global mosaics representing the passbands of each of the HRSC image filters (Neukum and Jaumann, 2004): panchromatic, infra-red, red, green, and blue, at present processed to 2 km/pixel and each in the form of six orthographic faces, with the pixel values of each calibrated to a single high quality observation area. While the pinning effect of the coverage gaps yields a map which appears partially detrended in comparison with TES albedo, and the absolute colour of any location should be interpreted with this in mind, the method offers the potential for obtaining filter-specific albedo maps without this caveat when we obtain complete coverage. During the mission extension until 2026 we hope to acquire a new observation encompassing the whole of the north polar cap near minimal ice cover as well as others to fill over the few remaining gaps or clouded areas, permitting an unpinned solution of the brightness models.

While it is unlikely that a four-filter colour composite mosaic will be used directly for analysis of surface composition on Mars, given the more capable spectral instruments already in use, the mosaic – and particularly the high resolution products derived from it – will find application in selecting areas of interest for study with those instruments, as well as in tracing out the extent of materials with identified spectral characteristics through continuity of colour over greater areas than could be achieved from spectral instruments alone. The colour-differentiating information expressed in composites of these filter mosaics is unprecedented for Mars, offering a remarkable new view of its surface, significant for geologic mapping and spectral correlation studies of composition as well as for the selection and planning of observations of new sites of particular compositional interest. The demonstration of colour-referenced high resolution images points to the possibility of future full-resolution colour maps from HRSC, and the methods should also find application in the processing of other (including historical) planetary image data and potentially spectral datasets, too.

**Data availability**

The mosaics can be downloaded from both the ESA PSA guest storage facility: https://doi.org/10.57780/esa-t4y39mj and from Freie Universität Berlin: http://dx.doi.org/10.17169/refubium-40624

**Acknowledgement**

We appreciate the comments and questions from anonymous reviewers which helped to clarify this manuscript. This work was supported by the German Space Agency (DLR Bonn), grant 50 OO 2204 (Koregistrierung), on behalf of the German Federal Ministry for Economic Affairs and Climate Action, and by the National Key Research and Development Program of China (grant 2022YFF0503100).